\documentclass[twocolumn,english,prl,superscriptaddress]{revtex4-2}
\usepackage{amsmath,amssymb,amsfonts}
\usepackage{CJK}
\usepackage{bm}
\usepackage{tikz}
\usepackage{pgfplots}
\usetikzlibrary{matrix,fit}
\usepackage{graphicx}
\usepackage{braket}
\usepackage{subfig}
\usepackage{caption}
\usepackage{mathrsfs}
\usepackage{float}
\usepackage[utf8]{inputenc}
\usepackage{pgflibraryarrows}
\usepackage{lipsum} 
\usepackage{color}
\usepackage{pgflibrarysnakes}

\begin{document}
\renewcommand{\raggedright}{\leftskip=0pt \rightskip=0pt plus 0cm}
\captionsetup[figure]{name={FIG.},labelsep=period}
\title{{Disentangled higher-orbital bands and chiral symmetric topology in confined Mie resonance photonic crystals}}

\author{Jing Li}
\affiliation{School of Science and Engineering, The Chinese University of Hong Kong, Shenzhen, 518172, China}

\author{Hongfei Wang}
\affiliation{Department of Physics, International Center for Quantum and Molecular Structures, Shanghai University, Shanghai 200444, China}

\author{Shiyin Jia}
\affiliation{School of Physics, Nanjing University, Nanjing, 210093, China}

\author{Peng Zhan}
\email[]{zhanpeng@nju.edu.cn}
\affiliation{School of Physics, Nanjing University, Nanjing, 210093, China}	

\author{Minghui Lu}
\email[]{luminghui@nju.edu.cn}
\affiliation{Department of Materials Science and Engineering, Nanjing University, Nanjing, 210093, China}

\author{Zhenlin Wang}
\affiliation{School of Physics, Nanjing University, Nanjing, 210093, China}	

\author{Yanfeng Chen}
\affiliation{Department of Materials Science and Engineering, Nanjing University, Nanjing, 210093, China}

\author{Bi-Ye Xie}
\email[]{xiebiye@cuhk.edu.cn}
\affiliation{School of Science and Engineering, The Chinese University of Hong Kong, Shenzhen, 518172, China}


\begin{abstract}
{Topological phases based on tight-binding models have been extensively studied in recent decades. By mimicking the linear combination of atomic orbitals in tight-binding models based on the evanescent couplings between resonators in classical waves, numerous experimental demonstrations of topological phases have been successfully conducted. However, in dielectric photonic crystals, the Mie resonances' states decay too slowly as $1/r$, leading to intrinsically different physics between tight-binding models and dielectric photonic crystals. Here, we propose a confined Mie resonance photonic crystal by embedding perfect electric conductors between dielectric rods, creating the chiral symmetric band structure which ideally matches tight-binding models with nearest-neighbour couplings. As a consequence, disentangled band structure spanned by higher atomic orbitals is observed. Moreover, our result provides an effective route to achieve a three-dimensional photonic crystal with complete photonic bandgap and third-order topology. Our implementation offers a versatile platform for study exotic higher-orbital bands and achieving tight-binding-like 3D topological photonic crystals.}

\end{abstract}
\maketitle

{\it{Introduction}}.---{The past decades has witnessed a fast development of topological physics aided by consecutive proposals of tight-binding models (TBMs) in electronic systems such as the Su-Schrieffer-Heeger (SSH) model\cite{SSH, SSH2, SSH3, SSH4}, the Kane-Mele model\cite{KM}, the Haldane model\cite{Haldane} and the Benalcazar-Bernevig-Hughes (BBH) model\cite{BBH}, etc. with distinct topological phases \cite{TI1, TI2, TI3}. Afterwards, direct analogies of these theoretical TBMs in artificial classical wave systems such as photonics, acoustics, mechanics, electric circuits, plasmonics, etc. lead to the successful observations of topological phases with intriguing physical effects in artificial materials\cite{PTI1, PTI2, PTI3, PTI4, PTI5, PTI6, PTI7, PTI8, d1, PNI1, PNI2, PNI3, PNI4, PNI5}. Especially, the discovery of topological phases in photonics has enabled many unprecedented optical effects and applications, such as unidirectional waveguides\cite{Waveguides}, topological cavities\cite{Cavity}, and topological lasers\cite{Laser}. A vast majority of photonic topological phases are based periodic array of dielectric rods (or drilling air holes in dielectric materials)\cite{PTI8, PTI9} which share a similar band structure and spatial symmetries to the corresponding TBMs. However, few attention has been paid to the discrepancy between the dielectric photonic crystals (DPCs) and TBMs. For example, at lower frequency, as opposed to the electronic TBMs, the host medium in dielectric photonic crystals supports propagating solutions for every frequency, leading to a linear dispersion around the Brillouin zone center. Besides the Mie resonances' states in DPCs are not exponential decay (confined states) in the rate of  $1/r$ when $r$ $\to$ $\infty$\cite{Mie}. These unique discrepancies will inevitably breaks the chiral symmetry of the photonic band structure, leading to the breaking of zero-energy character of the higher-order topological boundary states and the topological bound states in the continuum (BICs)\cite{BIC1, BIC2, BIC3}.} Moreover, in three-dimensional (3D) photonic crystals, the electromagnetic fields are generally vector fields (cannot be simply classified as transverse magnetic (TM) or transverse electric (TE) modes) and therefore the band structures are significantly different from those of TBMs. Achieving tight-binding-like photonic crystals in both 2D and 3D for all the bands still remains challenging.

Besides, TBMs have an important feature that stems from their linear combination of atomic orbitals (LCAO)\cite{orbital}: the local orbitals can be $s$-, $p$-, $d$-, $f$- and higher-orbitals, leading to $s$-, $p$-, $d$-, $f$- and higher-orbital bands. These different-order orbital band structures are critical for many significant physical phenomena, such as the existence of Majorana bound states in $p$-wave superconductors\cite{majorana}, the higher-temperature superconducting phases in $d$-wave superconductors\cite{Dwave} and multiple sets of corner states induced by the $d$-orbital bands\cite{Mcorner}. Most studies on photonic crystals have focused on the $s$-wave band structure\cite{DHOTI1}, and there have been a few examples studying the higher-orbital bands structure, which have shown intriguing physical effects. For instance, negative coupling can be realized in a photonic waveguide array with both $s$- and $p$- orbitals and that octupole insulators can be observed in $d$-wave photonic bands\cite{MPTI1, MPTI2, MPTI3, MPTI4}. However, these few works on the higher-orbital bands in photonic systems are limited in waveguide arrays which have a longitudinal direction and therefore possess a relatively large volume size\cite{MPTI1,MPTI} and can hardly be extended to 3D lattices. More importantly, the higher-orbital bands are always entangled and therefore toxic for applications. So far, to best of our knowledge, there is no effective way to systematically disentangling the higher-orbital band structures (denoted as disentangled higher-orbital bands). Thus, the exploration of various topological phases in higher-orbital bands is still very limited.

In this paper, we propose a new type of photonic crystal called confined Mie resonance photonic crystals (CMR-PCs) in both of 2D and even in 3D\cite{HOTI1, HOTI2, HOTI3, HOTI4, HOTI5, HOTI6}. By embedding metallic rods between the dielectric rods of original photonic crystals, CMR-PCs confine the slowly decaying Mie resonances’ states, leading to the same band structures as those of TBMs. Specifically, we achieve disentangled higher-orbital bands with chiral symmetry. Furthermore, we design a 3D complete bandgap with third-order topology in photonic crystals. Our work provides an unprecedented paradigm for studying disentangled higher-orbital bands and 3D topological complete bandgaps in PCs. Additionally, it opens up a new direction for exploring exotic physical phenomena arising from complicated higher-orbital coupling in lattices.

\begin{figure}
\centering
\includegraphics[scale=0.07]{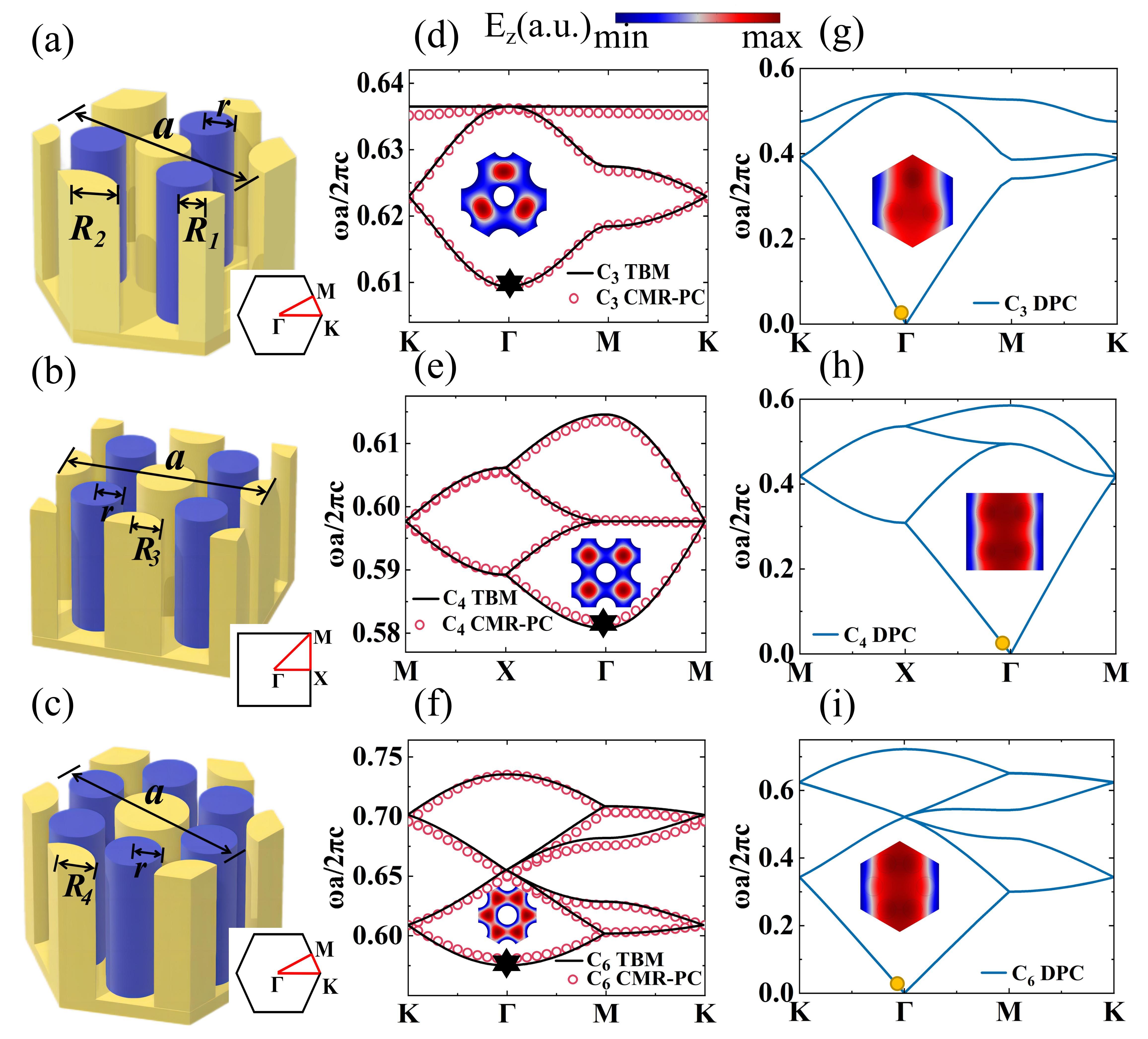}
\captionsetup{format=plain, justification=raggedright}
\caption{The structure of a (a) $C_3$- (b) $C_4$- (c) $C_6$- symmetric lattice (The yellow rods represent the metal and the blue rods represent the dielectric materials). The CMR-PCs'(red circles) and tight-binding models'(black lines) band structures with (d) $C_3$-, (e) $C_4$-, and (f) $C_6$- symmetric. The insets are field distributions of eigenmodes of photonic bands of the lowest frequency (Black stars). (g)-(h)-(i) Bands of (g) $C_3$-, (h) $C_4$-, and (i) $C_6$- symmetric DPCs and field distributions at the lowest frequencies respectively. (Orange circles).} 
\label{fig:1}
\end{figure}

{\it{Chiral symmetric photonic crystals}}.---In quantum mechanics, the Hamiltonian $\widehat{H}$ with chiral symmetry in momentum space is defined as $\widehat{U}\widehat{H}\widehat{U}(k)^\dag=-\widehat{H}(k)$, which implys the band structure is symmetric with respect to the zero energy. The chiral symmetry is closely related to many important physical properties. For example, the zero energy boundary states and the higher-order topological BICs is protect by the chiral symmetry in 1D  and 2D SSH model respectively\cite{BIC3}. However, in DPC, the chiral symmetry is intrinsically broken as the bands of DPC are linearly dispersive and the photons behave like plane wave in low frequencies in the structure.

\begin{figure*}[htbp]
\centering
\includegraphics[scale=0.1]{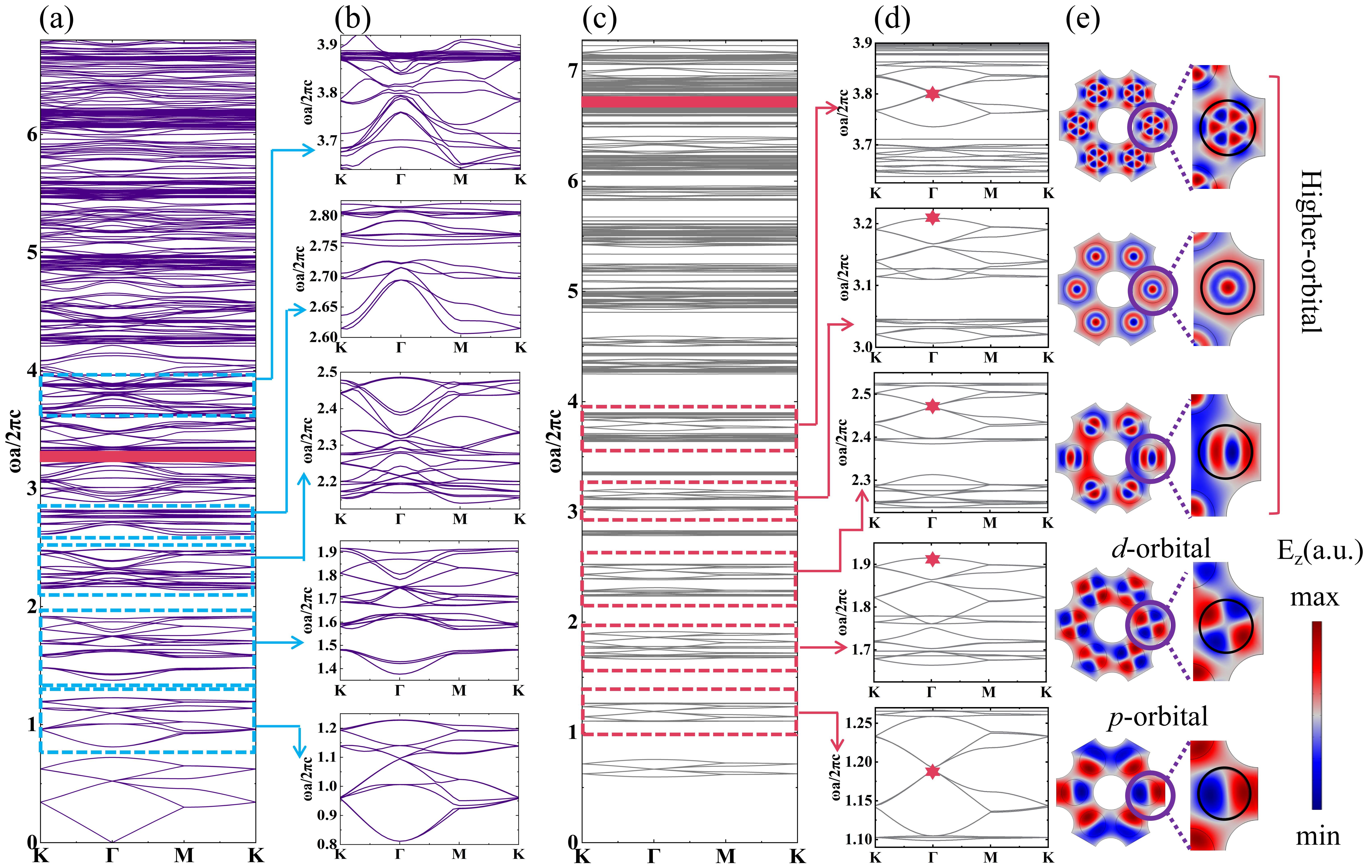}
\captionsetup{format=plain, justification=raggedright}
\caption{(a) The band structure of the DPC in Fig. 1(i) and the red area represents the highest bandgap. (b) Zoom-in pictures of the dashed box of (a) at normalized frequency 0.81-1.22, 1.37-1.92, 2.16-2.48, 2.61-3.23, and 3.64-3.92 respectively. (c) The band structure of CMR-PC in Fig. 1(c) and the red area represents the highest bandgap. (d) Zoom-in pictures of the dashed box of (c) at normalized frequency 1.09-1.27, 1.63-1.95, 2.23-2.55, 3-3.23, and 3.63-3.9 respectively. (e) Eigenmodes of different bands (red stars in (d)). The zoom-in pictures of (e) illustrate the higher-orbitals wave-function.}
\label{fig:2}
\end{figure*} 
 
To overcome this problem, we propose a CMR-PC to successfully achieve chiral symmetry in the band structure. We systematically study the 2D lattices with $C_3$, $C_4$, and $C_6$ rotational symmetries. The CMR-PCs are designed by embedding perfect electric conductors (PEC) in DPC (for example, we can use metal rods instead of PEC at mircowave frequencies), as shown in Fig. 1(a), (b) and (c). The parameters such as lattice constant, radius of crystal dielectric rods, and the relative dielectric permittivity of dielectric rods for all cases as $a=1.5$cm, $r=0.12a$, $\epsilon=9$ respectively, radius of metal rods for $C_3$-, $C_4$-, $C_6$-CMR-PC as $R_1=0.12a$, $R_2=0.2a$, $R_3=0.14a$, and $R_4=0.16a$.

Next, we calculate the bands and field distribution for TM modes. We find that the first and second bands of $C_3$-CMR-PC is symmetric with respect to central frequency(denoted as generalized chiral symmetry in Ref. \cite{GCS}), and the bands of $C_4$- and $C_6$-CMR-PC are symmetric with respect to their central frequencies (denoted as chiral symmetric bands). Additionally, these bands correspond well with the TBMs bands (black lines), as shown in Fig. 1(d), (e), (f). (See details of Hamiltonian for the 2D $C_3$, $C_4$  and $C_6$ SSH model in Section I of supplementary information\cite{SUP}.) Furthermore, we find that the $E_z$ components of the electromagnetic waves are localized around the single dielectric rods of CMR-PC even at the lowest frequency (Black star), resembling the LCAO in TBMs. On the contrary, when we remove metal rods, the low-frequency band (Orange circles) near the $\Gamma$ point tends is linearly dispersive and electromagnetic waves propagating in the structure have the form of plane waves (See Fig. 1(g), (h), (i)). Additionally, we find that the recovery of chiral symmetry of band successfully achieve BICs which have high Q-factors. (See higher-order topological BICs in Section II of supplementary information\cite{SUP}.)

{\it{Disentangled higher-orbital band}}.---The higher-orbital structures possess rich and intriguing physical phenomena, such as the orbital wavefunction of the surface states in 3D topological insulators \cite{HBTPC2}, topological states of ladder-like optical lattice high-orbital bands\cite{HBTPC3}, $p$-orbital disclination states\cite{HBTPC4}, etc. However, in photonics, the relevant higher-orbital structures is seldom been explored because of entangled higher-orbital bands. 

Here, in our CMR-PCs, the existence of chiral symmetry leads to an exact correspondence between photonic bands and band structures of TBMs. Therefore at high energy level with higher-orbital degree of freedom, the photonic band structure possesses disentangled dispersion of bands and band degeneracies at $\Gamma$ point. To demonstrate this feature, we calculate 400 bands for both the $C_6$-DPC and CMR-PC, we observe that the higher-orbital bands of DPC are completely entangled and for normalized frequencies above 3.3 (red area at Fig. 2(a)) the bands are close to a continuous domain, as shown in Fig. 2(a) and (b), which pose a great challenge for us to study the higher-orbitals coupling. However, we find that the higher-orbital bands are disentangled and still have a complete bandgap at the normalized frequency of 6.77 (red area at Fig. 2(c)) when we embed metal rods in DPC, as shown in Fig. 2(c). Furthermore, we find that higher-orbital bands maintain chiral symmetry and have degeneracy points at $\Gamma$ point, as shown in Fig. 3(d). Moreover, from the field distributions in one unit cell of modes in different bands (red stars in Fig. 3(d)), we observe a $p$-, $d$- and hybridized higher-orbitals degree of freedoms in rods, as shown in Fig. 3(e). The higher-orbital bands of $C_3$- and $C_4$-CMR-PC still have the same disentangled characteristic as the higher-orbital bands of $C_6$-CMR-PC. (See higher-orbital bands of $C_3$- and $C_4$-CMR-PC in Section III of supplementary information\cite{SUP}.)

\begin{figure}
\centering
\hspace{-0.5cm}
\includegraphics[scale=0.065]{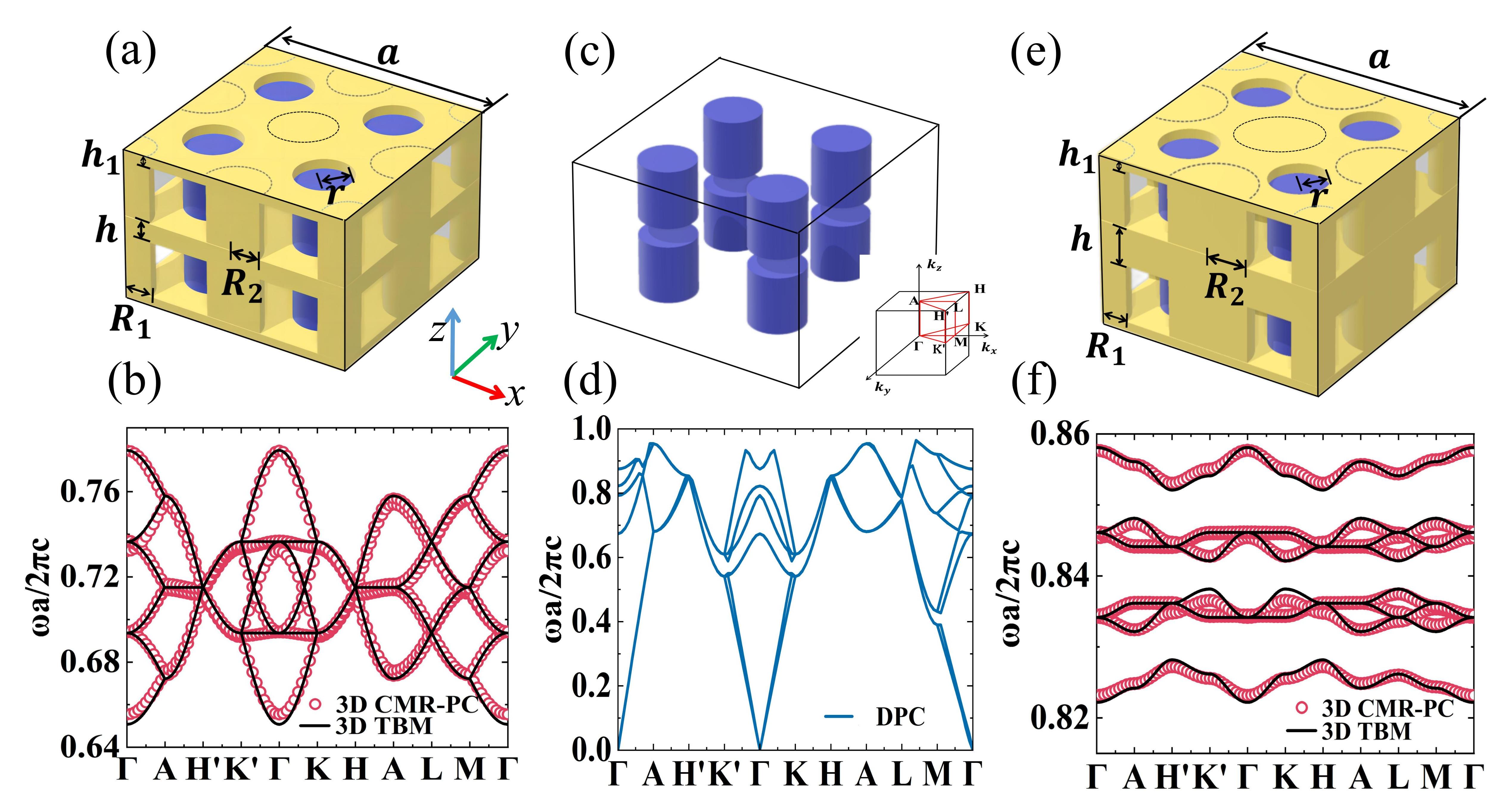}
\captionsetup{format=plain, justification=raggedright}
\caption{(a) The unit cell for a 3D CMR-PC. The yellow area represents the metal, the blue rods represent the dielectric material and the dotted lines represent positions of metal rods inside the layer. (b) Comparison of simulated bands for the 3D CMR-PC in (a) (red circles) with bands of corresponding TBMs (black lines). (c) The unit cell for a 3D DPC. (d) The Band structure for the 3D DPC in (c). (e) The unit cell for a 3D CMR-PC with non-trivial topology. (f) Comparison of simulated bands for the 3D CMR-PC in (e) (red circles) with the bands of corresponding TBMs (black lines).}
\label{fig:3}
\end{figure} 

{\it{3D complete bandgaps with third-order topology}}.---Currently, photonic higher-order topological phases is mostly studied in 2D lattices. Achieving a complete bandgap for 3D photonic crystals is important and challenging \cite{3PC1, 3PC2}. Here based on CMR-PCs, we design a 3D photonic crystal with a complete bandgap with third-order topology. The design of the 3D PC is achieved by stacking the 2D $C_4$-symmetric CMR-PC as shown in Fig. 1(b) in the vertical $z$ direction. The metal plates with air holes is used to control the coupling in the $z$ direction. Since the in-plane coupling is modulated by changing the radius of metal rods, the coupling in $z$ direction is not influenced by the changing of couplings in in-plane directions and can be controlled independently. Based on this design, we can simply take the $E_z$ mode into consideration and the electromagnetic waves in 3D photonic crystals are scalar. Therefore 3D band structures now exactly equal to corresponding tight-binding models, even at the lowest frequency around the Brillouin zone center.

We here consider three different types of 3D PCs. For all cases, the relative dielectric constant of the dielectric rods of model $\epsilon=9$. In Fig. 3(a), we design a structure with equal intercell and intracell couplings in all three directions by setting $a=1.5cm$,  $r=R_1=R_2=0.12a$, $h=a/12$, and $h_1=a/24$. (The higher-orbital bands of 3D CMR-PC exhibit similar disentanglement, see Section IV of supplementary information\cite{SUP}.) We calculate the bands of 3D CMR-PC and TBMs and find the band structures for Fig. 3(a) are symmetric with respect to central frequency (red circles) are perfectly matched to the TBMs band (black lines) with equal intercell and intracell couplings in three directions, as shown in Fig. 3(b). (See details of Hamiltonian of 3D SSH model in Section IV of SI\cite{SUP}.) In Fig. 3(c), we remove all the metal and keep other parameters the same as Fig. 3(a) (Momentum space and high symmetric points in the Brillouin zone (bottom right corner of Fig. 3(c)). However, we observe that the chiral symmetry of the bands is broken as shown in Fig. 3(d). Finally, to realize 3D topology, we adjust the intracell and intercell coupling by modifying the radius of the metal rods within layer and the height of the interlayer metal plates, as shown in Fig. 3(e). In Fig. 3(e), we design a 3D topological CMR-PC with complete bandgaps by setting $R_1=0.12a$, $R_2=0.18a$, $h=a/6$, and $h_1=a/24$ and set other parameters the same as Fig. 3(a). We see a large 3D complete bandgaps and a similarity to the bands of TBMs with intracell coupling: intercell coupling 1:6.6 (black lines) (See Fig. 3(f)). Moreover, we find that the Wannier center shift from the center of the unit cell through calculations of the Wannier center polarization of 3D topological CMR-PC. (See topological invariant calculation in Section V of supplementary information\cite{SUP}.)

To unveil the third-order topology, we design a 3D finite structure with $4\times4\times4$ topological unit cells in the Fig. 4(a). The topological properties are characterized by the Wannier center in unit cell \cite{3DHOTI1}. We calculate the eigenmodes of the finite-size structure and find that at 16.523 GHz, the field is distributed as bulk region which implys a bulk states (Fig. 4(b)). Then at 16.63 GHz, we find the field are strongly localized at the surface of the sample, representing a topological surface states and a first-order topology (see Fig. 4(c)). Besides, at 16.636 GHz, the field is localized at 1D hinges, demonstrating a topological hinge states and second-order topology (see Fig. 4(d)). Finally, at 16.741 GHz, a strong localization of fields at corner is observed, clearly revealing a corner states and third-order topology (see Fig. 4(e)). (Third-order topological SSH are discussed in Section VI of supplementary information\cite{SUP}).

\begin{figure}
\centering
\includegraphics[scale=0.055]{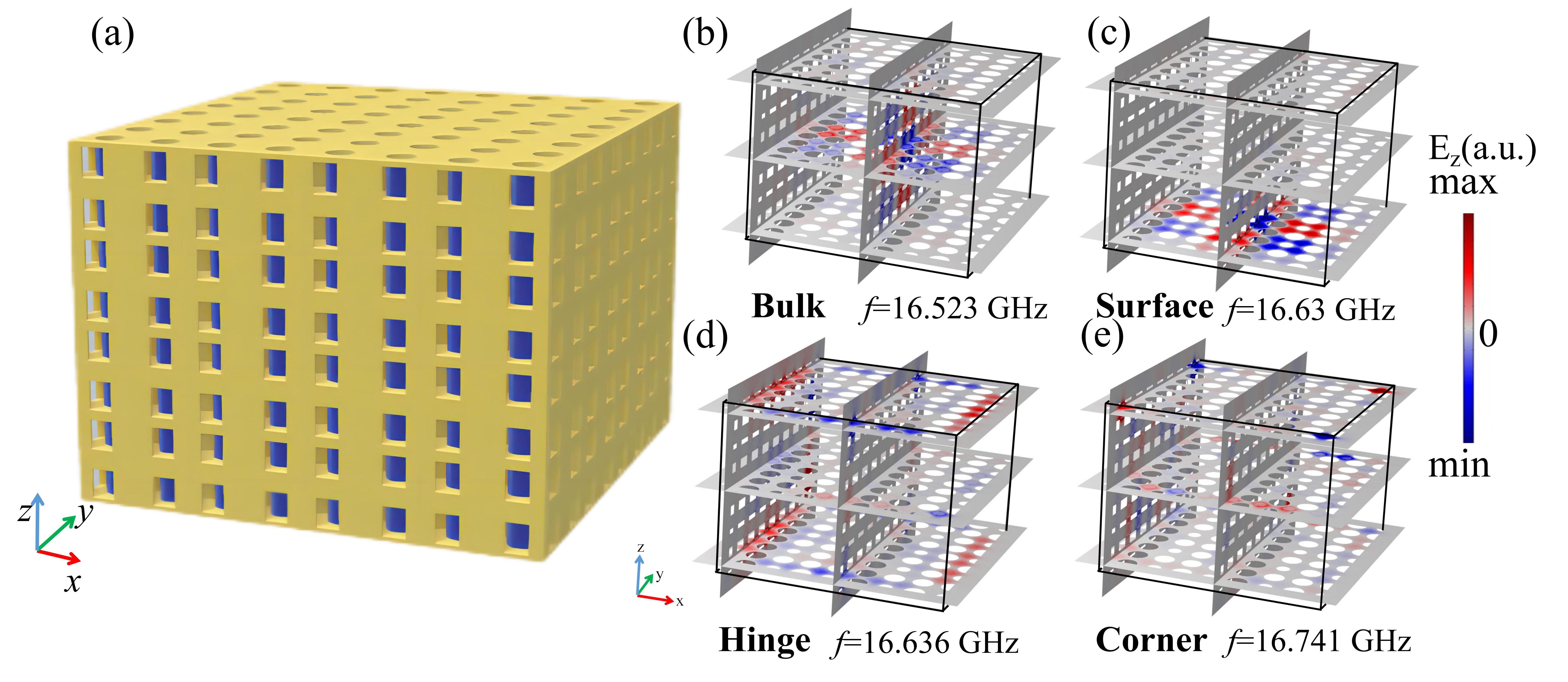}
\captionsetup{format=plain, justification=raggedright}
\caption{(a) 3D finite-size structure consists of  $4\times4\times4$ unit cells of CMR-PCs with non-trivial topology. (b) Bulk state at $16.523$ GHz. (c) Surface state at $16.63$ GHz. (d) Hinge state at $16.636$ GHz. (e) Corner state at $16.741$ GHz. Here black frame of (b)-(c)-(d)-(e) represents the outline of the finite-size structure.}
\label{fig:4}
\end{figure} 

{\it{Conclusion and disscussion}}.---In summary, we propose the CMR-PC as an ideal platform for studying chiral symmetric band structures and disentangled higher-orbital bands in photonic crystals. Our results open a convenient pathway to achieve topological BICs\cite{BIC1, BIC2}, higher-orbital bands and 3D complete bandgaps in photonic crystals. Intriguingly, as CMR-PCs have the same band structure as the TBMs, by breaking different space group symmetries, we may achieve various types of topological phases in 2D and 3D photonic crystals. Besides, if we further introduce gyromagnetic materials into the CMR-PCs, we expect to observe 3D higher-order Chern insulators with chiral hinge states. In addition, the exotic couplings between higher-orbital modes between dielectric rods may lead to many intriguing physical phenomena such as the negative couplings in photonic crystals with projective symmetry\cite{MPTI1, PFTBM1, PFTBM2}. Since the  band structures of CMR-PCs is tight-binding-like, we may further explore the 3D parity-time ($PT$) symmetric optics~\cite{PT}, the 3D topological defects \cite{defect}, 3D monopole topological resonators~\cite{luling 3d}, etc. in photonic crystals.

\begin{acknowledgments}
The authors acknowledge Xunqi Zhan and Zemeng Lin for useful discussion. This work was financially supported by National Key R$\&$D Program of China (Grants No. 2022YFA1404302 and No. 2018YFA0306200), National Natural Science Foundation of China (Grants No. 12174189 and No. 11834007), Stable Support Program for Higher Education Institutions of Shenzhen (Grants No. 20220817185604001), and the start-up funding at the Chinese University of Hong Kong, Shenzhen (UDF01002563).
\end{acknowledgments}

\end{document}